\documentclass[prb,twocolumn,notitlepage,longbibliography]{revtex4-2}
\usepackage{amsmath}
\usepackage{amssymb}
\usepackage{bbold} 
\usepackage{natbib}
\usepackage{lipsum}
\usepackage[unicode=true,colorlinks=true,citecolor=blue,urlcolor=blue]{hyperref}

\usepackage{bm}
\usepackage{epsfig} 
\usepackage{tikz} 
\usepackage[normalem]{ulem}

\renewcommand {\Im}{\mathop\mathrm{Im}\nolimits}

\renewcommand {\phi}{{\varphi}}
\newcommand {\rmi}{{\rm i}}

\newcommand {\e}{{\rm e}}

\begin{document}
\title{Emergent spin and orbital angular momentum of light in twisted photonic bilayer}
 % stacks of 2D layers

\author{Egor S. Vyatkin}
\affiliation{Ioffe Institute, St. Petersburg 194021, Russia}

\author{Alexander V. Poshakinskiy}
\affiliation{Ioffe Institute, St. Petersburg 194021, Russia}

\author{Sergey A. Tarasenko}
\affiliation{Ioffe Institute, St. Petersburg 194021, Russia}

\email{}

\begin{abstract}
We demonstrate that the optical response of twisted photonic bilayers, photonic counterparts of van der Waals structures, is sensitive to both spin angular momentum (SAM) and orbital angular momentum (OAM) of light. A beam of unpolarized light with zero angular momentum acquires SAM in transmission and OAM in reflection. The developed analytical theory and numerical calculations show that
the SAM and OAM arise from distinct microscopic mechanisms and depend differently on the interlayer distance. The predicted phenomena do not require light absorption and are caused by the photon-helicity-dependent light diffraction by the moiré pattern, which inevitably occurs in the twisted structure, and the SAM-OAM conversion processes. We also reveal strong SAM and OAM in the moiré-diffracted beams. Our findings uncover a profound connection between the emergent SAM and OAM in twisted photonic systems offering new possibilities for angular-momentum-resolved light-matter interactions.
\end{abstract}
\date{\today}

\maketitle
%%%%%%%%%%%%%%%%%%%%%%%%%%%%%%%%%%%%

\section{Introduction}\label{sec:intro}

Chiral objects, which are different from their mirror image, are famous for their ability to interact differently with photons of right-handed and left-handed circular polarizations, i.e., photons with the opposite projections of spin angular momentum (SAM).
This results in circular birefringence (CB) and circular dichroism (CD) manifesting, respectively, as the rotation of the polarization plane of linearly polarized light passing through the chiral medium and the emergence of a partial circular polarization for initially unpolarized beam~\cite{Fresnel1866,Agranovich1984}.
Recently, it became acknowledged that, in addition to SAM, photons can also possess orbital angular momentum (OAM) associated with the winding number of the wave phase~\cite{Bliokh2015,Knyazev2018} and chiral structures can discriminate beams with opposite OAM projections giving rise to optical vortex (or helical) birefringence and vortex dichroism (VD)~\cite{Babiker2018,Forbes2021rev}.
 For chiral molecules, which are much smaller than the light wavelength, VD becomes possible only beyond the electro-dipole electron-photon interaction~\cite{Andrews2004,Forbes2018} or beyond the paraxial approximation~\cite{Forbes2021}. Similarly to CD, VD can be used to distinguish 
enantiomers while providing extra sensitivity to the quadrupole transitions~\cite{Brullot2016}.
VD is also demonstrated for wavelength-scale particles made of chiral media~\cite{Wang2023,Cui2021} or asymmetrically-shaped objects 
made of achiral media~\cite{Woniak2019,Ni2021}. Furthermore, the Raman scattering by chiral molecules can be sensitive to the OAM of light~\cite{Forbes2019prl,Mullner2022} and, conversely, a beam with OAM can induce chirality in structures with 
nonlinear susceptibility~\cite{Begin2023,Nikitina2024}.  

\begin{figure}
    \centering
   \includegraphics[width=0.99\columnwidth]{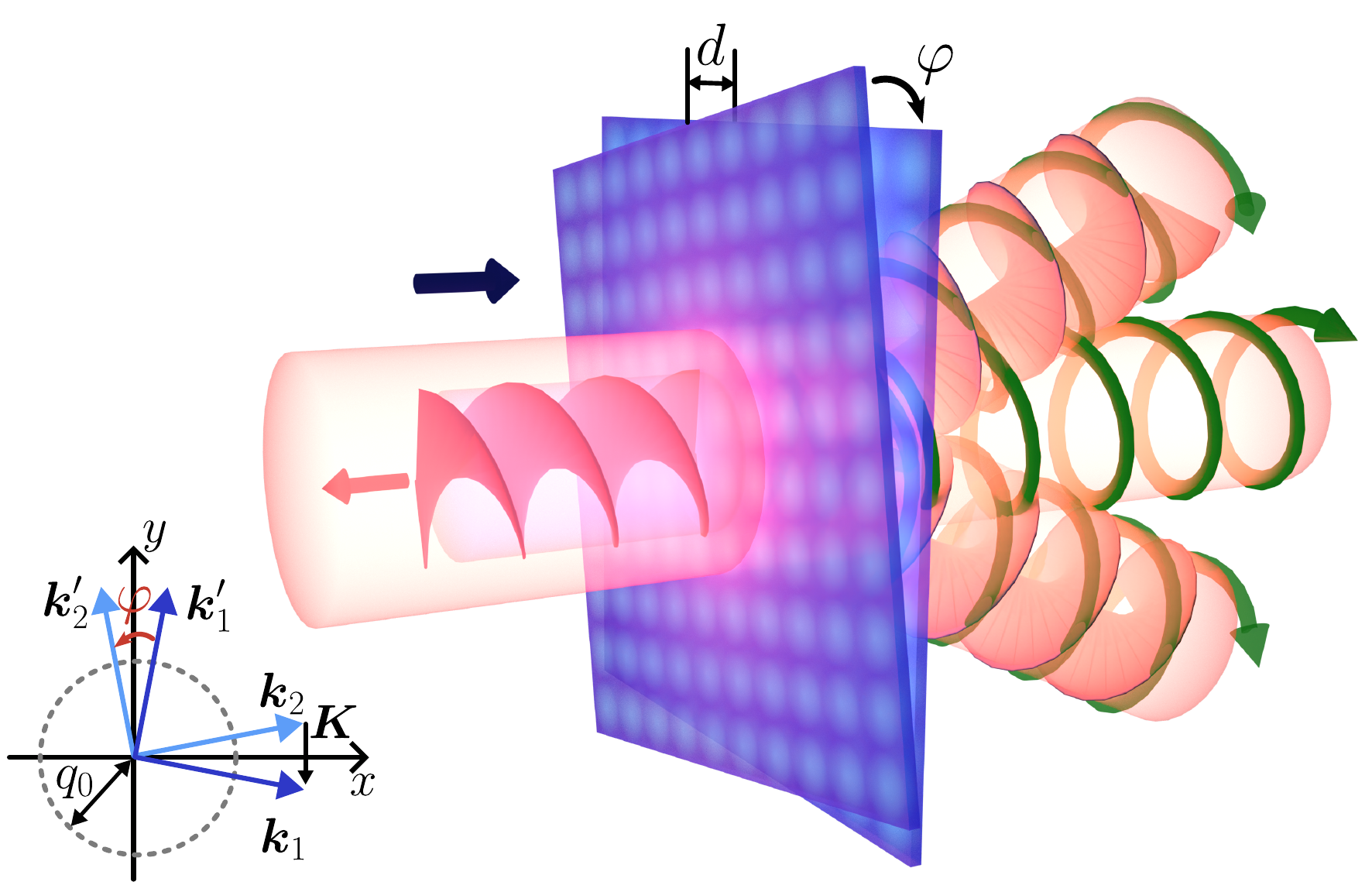}
    \caption{Sketch of twisted bilayer photonic crystal illuminated with normally incident beam of unpolarized light. The transmitted and reflected beams acquire spin angular momentum and orbital angular momentum, respectively. The beams diffracted by moir\'{e} pattern acquire both spin and orbital angular momenta. Insert shows the reciprocal lattice vectors of the first ($\bm k_1$, $\bm k_1'$) and the second ($\bm k_2$, $\bm k_2'$) layers yielding the first-order moir\'{e} diffraction vectors $\bm K = \bm k_1-\bm k_2$ and $\bm K' = \bm k'_1-\bm k'_2$.}
    \label{fig:scheme}
\end{figure}

To enable chiral response in solid state optics, planar waveguides and microcavities with lithographically patterned chiral elements are proposed~\cite{Konishi2011,Maksimov2014}. 
Another approach to create a solid state chiral metasurface is to take a pair of achiral photonic layers and stack them with a twist~\cite{Han2022}, 
as shown in Fig.~\ref{fig:scheme}. Such a design provides additional tunabilty through the control of the twist angle $\varphi$ and the interlayer distance $d$.
It is inspired by twisted van der Waals crystals, such as bilayer graphene, where the fine tuning of the twist angle enables the emergence of 
 electronic flat bands and superconductivity~\cite{Cao2018,Torma2022}, and twisted TMDC heterostructures exhibiting intrinsic circularly polarized exciton emission~\cite{Michl2022}.
 Twisted  bilayers of optical lattices for cold atoms~\cite{Gonzalez2019,Meng2023}, mechanical crystals~\cite{Deng2020,Lopez2020}, and nodal superconductors~\cite{Volkov2023} are also considered. In twisted photonic lattices, light localization due to the optical flat bands~\cite{Wang2019}, formation of solitons~\cite{Fu2020}, the edge, corner, and other twist-dependent optical modes~\cite{Lu2023,Arkhipova2023,Salakhova2023} are reported. However, the aforementioned effects are caused by the in-plane moir\'{e} modulation and do not require the chirality per se. The study of chiral optical response is limited so far to CD and CB, which was demonstrated for twisted bilayer graphene~\cite{Kim2016} and twisted stacks of other 2D semiconductors~\cite{Poshakinskiy2018} as well as their optical counterparts~\cite{Lou2021,Lou2023}.

In this paper, we show that twisted photonic bilayers exhibit both SAM- and OAM-dependent response. In particular, when the bilayer is illuminated by a beam of unpolarized light with zero angular momentum, the transmitted and reflected beams acquire SAM and OAM. While the two phenomena have common phenomenological origin in the chirality of the system, we reveal that they have different microscopic nature, which leads to their very distinct properties. (i) SAM originates from the near-field interaction between the layers and decays with the interlayer distance while OAM comes from the interference of the SAM-OAM conversion in the layers and
oscillates with the interlayer distance with no decay. (ii) SAM is observed in transmission while OAM is maximal in reflection. Moreover, we show that both phenomena do not require light absorption and are observed also in transparent bilayers. The emergence of SAM in the transmitted beam is caused by the 
diffraction of light by the moir\'{e} pattern, which inevitably occurs in the twisted structure and appears to be highly sensitive 
to the helicity of incident photons. We also study the parameters of the moir\'{e} diffracted beams and show that the beams
possess significant SAM and OAM.

\section{Model and Method}\label{sec:model}

We consider a chiral bilayer consisting of two ($j=1,2$) thin dielectric slabs, Fig.~\ref{fig:scheme}. 
The layers are positioned at $z_1  = 0$, $z_2=d$ and rotated relatively to each other by the angle $\varphi$.
Each layer is characterized by the 2D permittivity
which is periodically modulated 
in the form of a square lattice with the period $2\pi/k$, so that the in-plane 2D polarizations  are given by 
\begin{align}\label{eq:Pj} 
    \boldsymbol{P}^{(j)}(\bm \rho)=&\left[\alpha_0+2\alpha_1(\cos \boldsymbol{k}_j\boldsymbol{\rho}+\cos \boldsymbol{k}_j'\boldsymbol{\rho})\right]\boldsymbol{E}_\parallel(\bm \rho ,z_j) \,,
\end{align} 
where $\bm \rho$ is the in-plane coordinate, 
$\alpha_0$ is the uniform component of the layer polarizability and $\alpha_1$ is the amplitude of its spatial modulation, 
$\bm k_j$ and $\bm k'_j$ are the reciprocal lattice vectors of the modulation, 
$|\bm k_j| = |\bm k'_j| = k$, $\bm k_j \perp \bm k_j'$, $\bm k_1 \cdot \bm k_2 = \bm k_1' \cdot \bm k_2' = k^2 \cos \varphi$, 
and $\boldsymbol{E}_\parallel(\bm \rho ,z_j)$ is the local in-plane electric field. The 2D polarizability depends on the real thickness of the dielectric slab.

The bilayer is illuminated by normally-incident monochromatic optical beam with the electric field 
$
\bm E_{\rm in}(\bm \rho ,z) = \int \bm E_{\rm in}(\bm q_{\parallel})\, \e^{\rmi (\bm q_\parallel \cdot \bm \rho + q_z z)} d\bm q_\parallel \,,
$
where $\bm E_{\rm in}(\bm q_{\parallel})$ is the in-plane Fourier image, $q_z = \sqrt{q_0^2-q_\parallel^2}$, $q_0=\omega/c$, and $\omega$ is the radiation frequency.
We consider Gaussian beams, $
\bm E_{\rm in}(\bm q_{\parallel}) \propto \e^{-a^2q_\parallel^2/2}\,  \bm e 
$
 with the width $a \gg 1/q_0$, so that the paraxial approximation is applicable and the electric field polarization $\bm e$ can be assumed constant. 
 
 Solution of the Maxwell equations yields the field distribution in the whole space in the form, see Appendix~\ref{app:sol},
\begin{equation}\label{eq:Eall}
\bm E(\bm \rho ,z) = \bm E_{\rm in}(\bm \rho ,z) + \int d \bm q_\parallel \sum_{\boldsymbol{g}}
e^{\rmi (\bm q_\parallel + \boldsymbol{g}) \cdot \boldsymbol{\rho}}  \bm E_{ \bm g}' (\bm q_\parallel ,z) \,,
\end{equation}
where $ \bm E_{ \bm g}' (\bm q_\parallel ,z)$ is the Fourier harmonic of the scattered field and the summation is performed over all diffraction wave vectors 
\begin{align}\label{eq:g}
\boldsymbol{g} = n_1 \boldsymbol{k}_1 + n_1' \boldsymbol{k}_1' + n_2 \boldsymbol{k}_2 + n_2' \boldsymbol{k}_2'
\end{align} 
with integer $n_j$ and $n_j'$. 
We focus on the ``metasurface'' regime when 
the period of permittivity modulation in the layers is smaller than the wave length of incident light ($k > q_0$). Then, individual layers do not produce propagating diffracted beams. However, in twisted bilayers, because of coupling via near fields,  the diffraction by the moir\'e pattern becomes possible, since its wave vector  $K = |\bm k_1-\bm k_2|$ can be smaller than $q_0$, see Fig.~\ref{fig:scheme}. 
In numerical solution, to smoothen the sharp resonances we use the regularization
$\omega \to \omega + \rmi \gamma$ with small $\gamma$ which corresponds to the medium with weak losses.

For the incident beam much wider than the  moir\'e scale ($1/a \ll K$), the diffracted beams of different orders do not overlap in $\bm q$-space.
Then, the Fourier components of the transmitted and reflected beams can be related to those of the incident beam via  the 2$\times$2 polarization Jones matrices $\bm t(\bm q_\parallel)$ and $\bm r(\bm q_\parallel)$. Similarly, we introduce the Jones matrices $\bm t^{(\bm g)}(\bm q_\parallel)$ and $\bm r^{(\bm g)}(\bm q_\parallel)$ for the diffracted beams with $\bm g \neq 0$ propagating forward and backward, respectively, see Appendix~\ref{app:sol}.

\begin{figure*}[t!]
    \centering
    \includegraphics[width=0.85\textwidth]{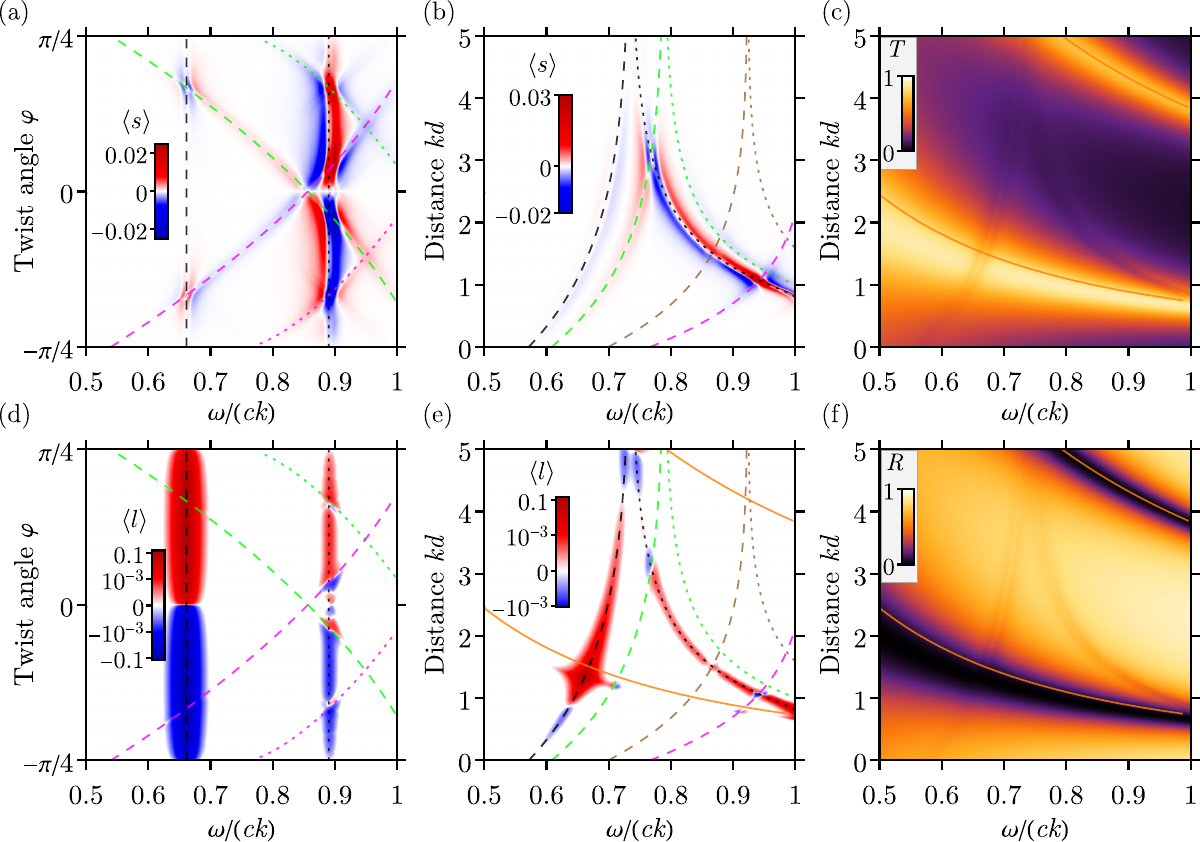}
    \caption{(a) SAM $\langle s \rangle$ of the transmitted beam 
 and (d) OAM  $\langle l \rangle$ of the reflected beam as a function of the light frequency $\omega$ and the twist angle $\varphi$ for the interlayer distance $kd=1.34$. (b) SAM of the transmitted beam and (c) transmittance $T$,
(e) OAM of the reflected beam and (f) reflectance $R$ as a function of the light frequency $\omega$ and the interlayer distance $d$ for the twist angle $\varphi=\pi/8$. Positions of the Fabry-P\'erot resonances and the resonances associated with symmetric and antisymmetric waveguide modes are shown by solid, dashed, and dotted lines, respectively. 
The figures are calculated for $\alpha_0 k = 0.2$, $\alpha_1 k = 0.01$,  $a k = 50$, and $\gamma /(ck) = 0.01$.} 
    \label{fig:Trans-and-Refl}
\end{figure*}

\section{Results}\label{sec:res}

We now discuss the properties of the transmitted, reflected, and diffracted beams. We characterize each beam by the intensity, the SAM $\langle  s \rangle$ (the degree of circular polarization), and the OAM $\langle l \rangle$. The latter is calculated using the operator 
$\bm l =- \rmi \bm n \cdot (\bm q \times \partial_{\bm q} )$ where $\bm n $ is unit vector in the  direction of beam propagation.

\subsection{General notes} 

Before presenting the results of calculations, we draw some general conclusions based on symmetry arguments. First, we note that individual photonic layers constituting the bilayer are described by the $D_{4h}$ point group, which is achiral and centrosymmetric. The transmission and reflection coefficients of individual layers are polarization independent for normally incident beams.
The bilayer becomes chiral if the layers are stacked with the twist providing the twist angle $\varphi \neq \pi n/4$, where $n$ is integer, and the interlayer distance $d \neq 0$. The symmetry of such a bilayer is reduced to the $D_4$ point group, which enables the emergence of SAM and OAM for normally incident beams.

Second, one can formulate optimal conditions on the interlayer distance $d$. It should not be too small since the bilayer lacks chirality at $d = 0$. On the other hand, it should not be too large otherwise the layers get optically decoupled. The interaction via near fields, essential for the emergence of SAM and moir\'{e} diffraction, is effective if the interlayer distance $d$ is comparable to the wavelength of permittivity modulation $2\pi /k$.
Assuming that $k$ is of the order of the photon wave vector $q_0$ we conclude that the optimal values of $d$ are in the $\mu$m scale for the visible spectral range and in the sub-mm scale for terahertz radiation.

Third, the $D_4$ point-group symmetry of the chiral bilayer together with the Lorentz reciprocity imposes the constraints on the transmission and reflection coefficients at $\bm q_\parallel =0$. We formulate them in the basis of circularly polarized plane waves ($\sigma=\pm 1$). At transmission, the sign of the circular polarization is conserved, $t_{\sigma,\sigma'}(0) = t_{\sigma} \delta_{\sigma, \sigma'}$. However, the coefficients $t_{+}$ and $t_{-}$ can be different, which enables the emergence of SAM $\langle s \rangle = (|t_+|^2-|t_-|^2)/(|t_+|^2+|t_-|^2)$ even for unpolarized incident beam. In reflection, $r_{\sigma,\sigma}(0) = 0$ and $r_{+,-}(0) = r_{-,+}(0) \equiv  r$ (we define the sign of the circular polarization with respect to the beam propagation direction), and the SAM $\langle s \rangle = (|r_+|^2-|r_-|^2)/(|r_+|^2+|r_-|^2)$ vanishes. Thus, the transmitted light acquires SAM  while the reflected light does not.

To understand the emergence of OAM one should consider optical beams constructed from the plane waves with nonzero $\bm q_\parallel$. At finite $\bm q_\parallel$, the above constraints on the transmission and reflection coefficients are violated. The SAM-OAM conversion processes take place such as $L_z = 0,\; S_z = \pm1 \rightarrow L_z =\pm 2,\;S_z = \mp 1$ and $L_z = 0,\; S_z = \pm1 \rightarrow L_z = \mp2 ,\;S_z = \mp1$, where $L_z$ and $S_z$ are the OAM and SAM projections on the incident beam axis. The former processes conserve the total photon angular momentum projection $L_z+S_z$.  
In the latter processes, the total photon angular momentum projection changes by $\pm4$ and the difference is converted into the mechanical rotation of the twisted bilayer.
They are allowed by 4-fold rotation symmetry of the bilayer and play a major role in the OAM emergence.
The difference in the rates of the conversion processes results in the OAM in the reflected beam even if the incident beam carries no OAM ($L_z = 0$) and consists of the equal portions of the photons with $S_z = \pm 1$.

\subsection{Transmitted and reflected beam}

First, we discuss the transmitted and reflected beams. We suppose that the incident beam is  unpolarized, i.e., consists of 
incoherent $\sigma^+$ and $\sigma^-$ contributions. Figure~\ref{fig:Trans-and-Refl} reveals that the transmitted beam acquires SAM 
[Figs.~\ref{fig:Trans-and-Refl}(a) and~\ref{fig:Trans-and-Refl}(b)] while the reflected beam acquires OAM [Figs.~\ref{fig:Trans-and-Refl}(d) and~\ref{fig:Trans-and-Refl}(e)]. Both effects are odd in the twist angle $\varphi$ and vanish at $\varphi = 0, \pm \pi/4$ when the bilayer lacks chirality, see Figs.~\ref{fig:Trans-and-Refl}(a) and~\ref{fig:Trans-and-Refl}(d). 
 The obtained values of the SAM $\langle s \rangle$ and the OAM $\langle l \rangle$ are non-integer. It means that the beams are the mixtures of the states with integer SAM and OAM projections with different portions. In our case, the mixture is formed by the states with $S_z = \pm 1$ and $L_z = 0, \pm2$.

The optical response of the bilayer contains the Fabry-P\'erot resonances (solid lines in Fig.~\ref{fig:Trans-and-Refl}) and the resonances associated with the guided modes confined between the two layers (dashed and dotted lines in Fig.~\ref{fig:Trans-and-Refl})~\cite{Lou2021}. The resonances are clearly seen in the transmittance $T$ and reflectance $R$, see Figs.~\ref{fig:Trans-and-Refl}(c) and~\ref{fig:Trans-and-Refl}(f), respectively. 
The Fabry-P\'erot interference leads to the oscillations in the transmittance and reflectance almost from 0 to 1 and is well described by the equations
\begin{equation}
    T=\frac{1}{|(1-2\pi\rmi \alpha_0 q_0)^2+(2\pi \alpha_0 q_0)^2 \e^{2\rmi q_0 d}|^2}, \;\; R=1-T ,
\end{equation}
which are derived neglecting the lateral modulation of the permittivity.
The resonances associated with the guided modes are much weaker in the transmittance and reflectance. They get visible
because the in-plane modulation of the permittivity $\alpha_1$ couples the guided modes to the free-space modes. The guided modes acquire finite lifetime and can be excited by normally incident beam. 
The frequencies of the symmetric and antisymmetric guided modes with the in-plane wave vector $\bm g$ and the Fabry-P\'erot resonances with $\bm g =0$ in the limit of small $\alpha_1$ are determined by the equation
\begin{equation}
    2\pi q_0^2 \alpha_0\left(1\pm \e^{-\sqrt{g^2 - q_0^2}d}\right)=\sqrt{g^2 - q_0^2} \,.
\end{equation}
Both SAM and OAM are enhanced at the resonances.
For our case of weak dielectric modulation $\alpha_1k\ll 1$, the enhancement is particularly pronounced at the frequencies of the guided modes with the first-order diffraction wave vectors $|\bm g| = k$ (black dashed and dotted lines in Fig.~\ref{fig:Trans-and-Refl}). 
At the resonances, the SAM changes sign as the function of the light frequency. This behavior is similar to the spectral dependence of the circular dichroism at the exciton resonance in bulk gyrotropic crystals~\cite{Agranovich1984}, quantum wells~\cite{Kotova2016}, and   
chiral stacks of 2D semiconductors~\cite{Poshakinskiy2018}.
Microscopically, the inversion originates from the splitting of the spectral positions of the resonances which are excited by $\sigma^+$ and $\sigma^-$ circularly polarized waves with the given wave vector.
The OAM of the reflected beam is additionally increased at the intersections of the guided modes with the Fabry-P\'erot modes, where the conventional reflection is suppressed. 

\begin{figure}[t!]
    \centering
    \includegraphics[width=0.99\columnwidth]{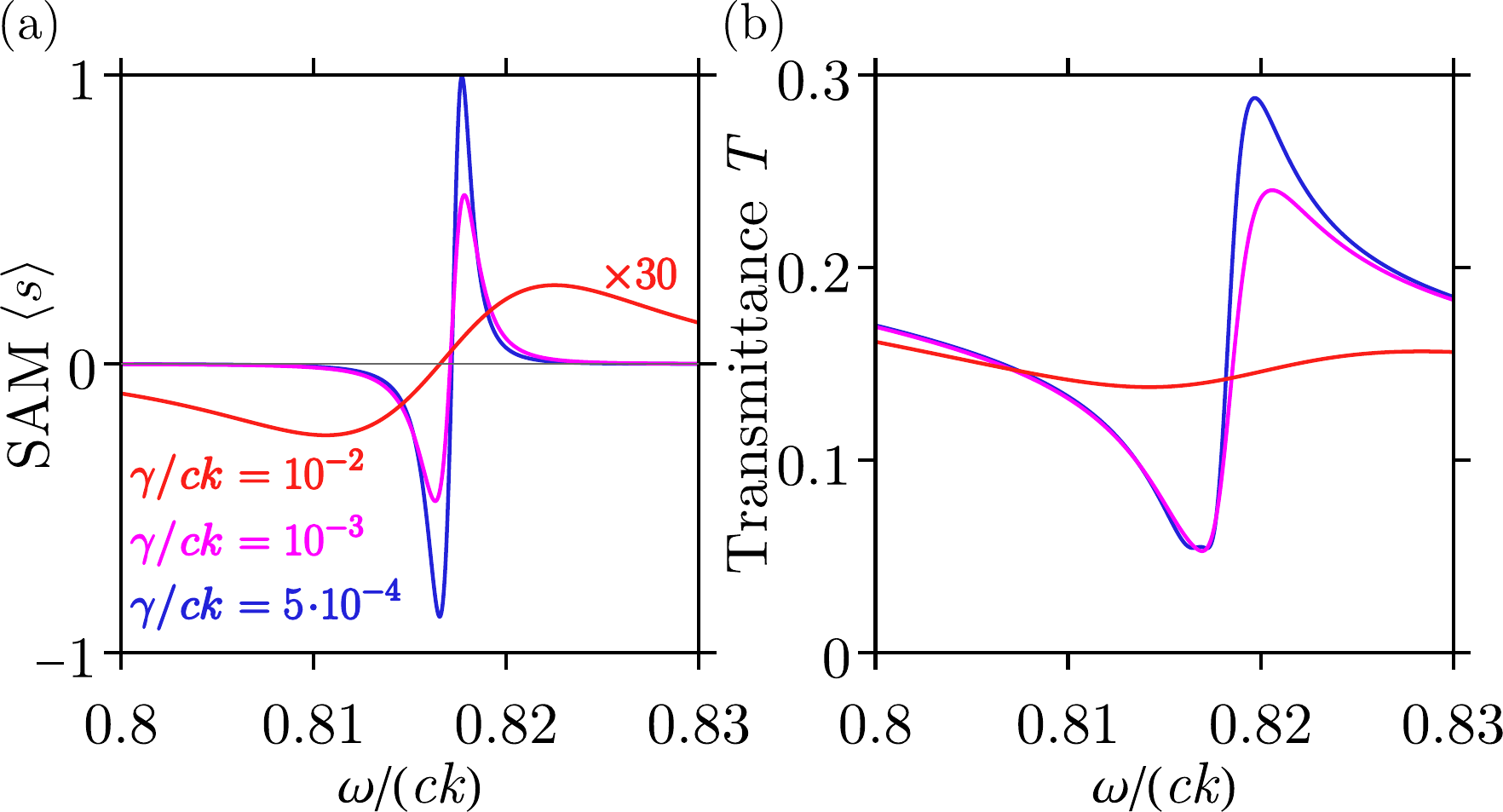}
    \caption{(a) SAM $\langle s \rangle$ of the transmitted beam and (b) bilayer transmittance $T$ as a function on the light frequency in the vicinity of a guided mode resonance. The curves are calculated for $kd = 2$, $\varphi = \pi/8$, and three different rates of energy losses $\gamma$. Other parameters are listed in the caption to Fig.~\ref{fig:Trans-and-Refl}.}
    \label{fig:SAMGamma}
\end{figure}

The amplitudes and widths of the guided mode resonances in Fig.~\ref{fig:Trans-and-Refl}
are determined by the background energy losses which are included in the numerical calculations. The used rate of the losses $\gamma = 10^{-2} ck$ is small but still exceeds the radiative broadening of the resonances determined by their coupling to propagating modes. Figure~\ref{fig:SAMGamma} shows the evolution of the SAM and 
the bilayer transmittance at a guided mode resonance with the decrease of the losses. The resonance gets much narrower and the SAM values reach $\pm1$. Further decrease of $\gamma$ does not affect much the results indicating that the limit of radiative broadening is almost achieved. Figure~\ref{fig:SAMGamma} reveals that
the fine tuning of the guided mode resonances to the desired light frequency
enables the achievement of high SAM in the transmitted light even in a single twisted bilayer.

Though the SAM (circular polarization) of the beam transmitted through the chiral bilayer may seem to be natural, 
its microscopic origin is in fact unusual and differs from the known mechanism. In conventional chiral media, the circular polarization of 
the transmitted (initially unpolarized) beam is caused by the circular dichroism, i.e., the difference in the absorption of $\sigma^+$ and $\sigma^-$ waves.  
Here, the circular polarization
emerges even in the absence of losses, which is particularly pronounced in Fig.~\ref{fig:SAMGamma}. 
Surprisingly, the transmission coefficients for $\sigma^+$ and $\sigma^-$ photons are different
although there is no absorption, the photons retain their helicity and the reflection coefficients for  $\sigma^+$ and $\sigma^-$ photons coincide.
This puzzle is solved by taking into account the channels of photon decay 
associated with the moir\'{e} diffraction. Indeed, twisting the bilayer produces 
the moir\'{e} pattern and unavoidable diffraction of light by the 
moir\'{e} wave vectors. In chiral bilayers, the intensity of the diffracted beams depends on the circular polarization of the 
incident beam. This difference in the decay of incident $\sigma^+$ and $\sigma^-$ photons over the moir\'{e} diffraction channels 
gives rise to the circular polarization of the transmitted beam.

Simple analytic expressions for the SAM and OAM can be obtained  in the perturbation theory in the case of small layer polarizability, $\alpha_0 k , \alpha_1 k \ll1$, see Appendix~\ref{app:pert}. The spin-dependent contribution to the transmission coefficient comes from the fourth-order diffraction processes, e.g., consequently by $\bm k_1$, $-\bm k_2$, $-\bm k_1$, $\bm k_2$. 
For small $\varphi$, the SAM of the transmitted beam assumes the form
\begin{align}\label{eq:st}
\langle s \rangle =
\frac{4\varphi q_0^2}{\varkappa^2} \, (2\pi \alpha_1 k)^4 \sin(2q_0 d) \,\e^{-2\varkappa d} \,,
\end{align}
where $\varkappa=\sqrt{k^2-q_0^2}$ is the reciprocal decay length of the near field. As the function of the interlayer distance $d$, the SAM rises linearly at small $d$, reaches maximum at $d \sim 1/\varkappa$, and  oscillates and decreases exponentially at large $d$.
Such a non-monotonous dependence also persists in the full calculation Fig.~\ref{fig:Trans-and-Refl}(b), where the SAM is shown as a function of the light frequency and the interlayer distance.
 The oscillations originate from the Fabry-P\'erot interference in the bilayer.
The exponential decrease at large $d$ clearly indicates that the SAM originates from the near-field coupling of the layers. Indeed, at large interlayer distances $d \gg 1/\varkappa$, the interaction of light with each layer can be considered independently. Since the individual layer possesses high symmetry ($D_{4h}$ point group), it exhibits polarization-independent transmission and reflection in the paraxial approximation, so that the SAM vanishes.

The OAM for the reflected beam arises already in the second-order diffraction processes, e.g., consequently by $\bm k_1$ and $-\bm k_1$. Such processes occur from  individual layers and lead to distortions of
the phase surface of the Gaussian beam. Interference of the distortions induced by the first and the second layers leads to the OAM 

\begin{align}\label{eq:lt}
    \langle l\rangle=-\frac{9\pi^2\alpha_1^4 k^8 }{2 \alpha_0^2a^4 \varkappa^{10} } \sin(4\varphi)\tan(q_0 d) \,.
\end{align}

\begin{figure}[t!]
    \centering
    \includegraphics[width=0.99\columnwidth]{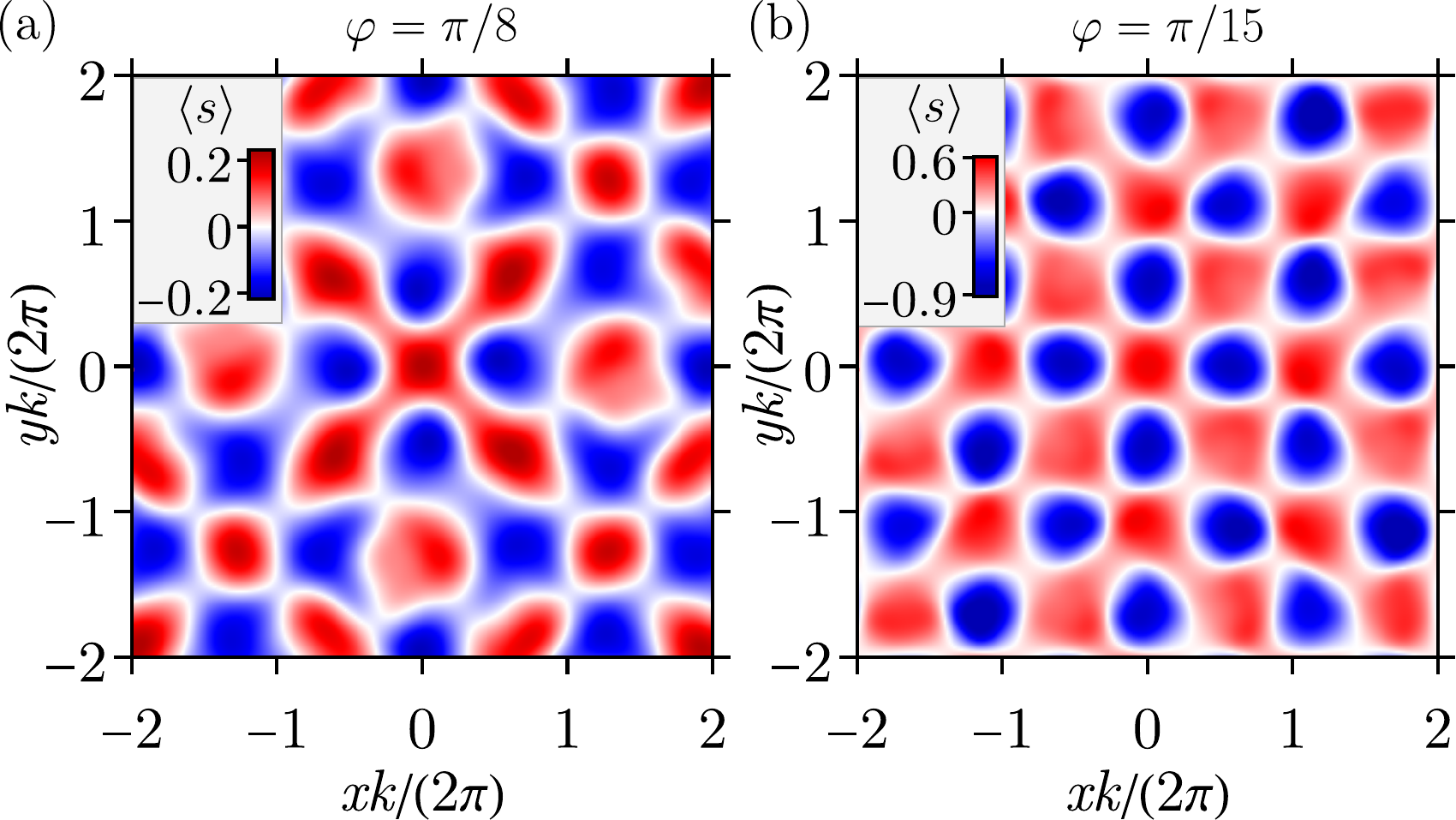}
    \caption{
    Spatial distribution of SAM in the plane $z=d/2$ between the layers 
    for the twist angle (a) $\varphi=\pi/8$ and (b) $\varphi=\pi/15$. The figures are calculated for $kd=2$, $\omega/(ck)=0.816$, and unpolarized incident light. 
    Other parameters are listed in the caption to Fig.~\ref{fig:Trans-and-Refl}.}
    \label{fig:Inside}
\end{figure}

Similarly to the SAM, the OAM vanishes at $\varphi=0,\pm \pi/4$ and reaches its maxima at the twist angle $\varphi = \pm \pi/8$.
The reflected beam OAM also oscillates as the function of the interlayer distance $d$, which originates from the Fabry-P\'erot interference.
However, being determined by the interference of the far fields rather than by the near fields, the OAM does not decay with the increase of the interlayer distance. 
In transmission, the optical paths of the beams distorted by the first and second layers are the same, suppressing their interference and leading to a much smaller value of the OAM than for the reflected beam.

In conclusion of this subsection, we comment that the near field in a twisted bilayer has a complex structure being the superposition of propagating modes and evanescence modes with different wave vectors $\bm g$. Figure~\ref{fig:Inside} shows the spatial distributions of the SAM in the plane $z = d/2$ centered between the layers. The distributions are calculated for the parameters $kd =2$ and $\omega/(ck)=0.816$ corresponding to a guided mode resonance, see Fig.~\ref{fig:Trans-and-Refl}(b), and two different twist angles $\varphi$. 
As expected, the SAM oscillates in the plane. The local values of the near field SAM are much higher than the SAM in the transmitted light, cf. Fig.~\ref{fig:Trans-and-Refl} and Fig.~\ref{fig:Inside}.

\subsection{Moir\'e diffracted beams}

\begin{figure}[t!]
    \centering
    \includegraphics[width=0.99\columnwidth]{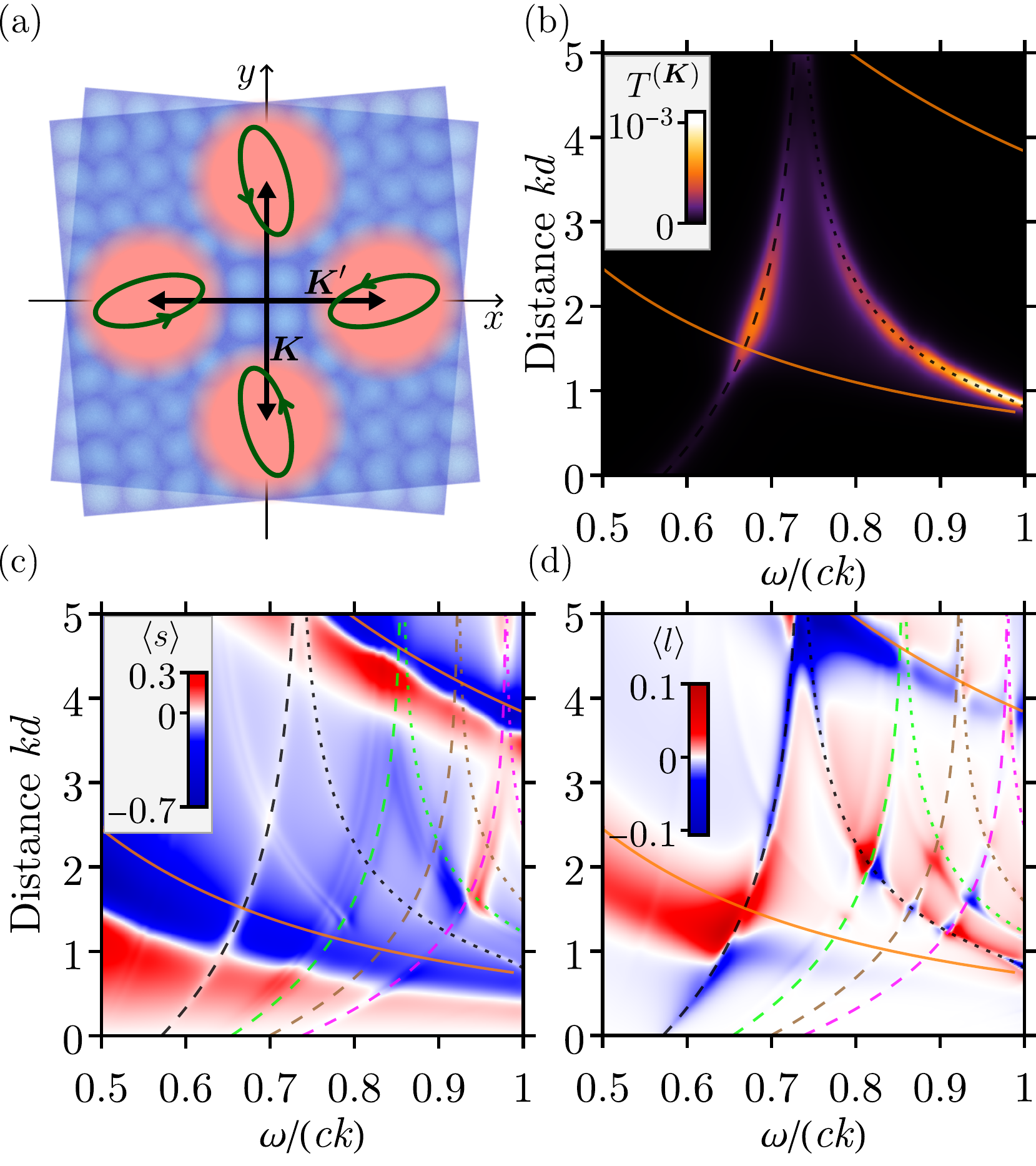}
    \caption{(a) Four moiré diffracted beams and their elliptical polarizations. (b) Intensity, (c) SAM  $\langle s \rangle$, and 
    (d) OAM $\langle l\rangle$ of the diffracted beam as a function of the interlayer distance $d$ and the light frequency $\omega$ for the twist angle $\varphi=\pi/15$. Other parameters are listed in the caption to Fig.~\ref{fig:Trans-and-Refl}.}
    \label{fig:Diff-Gauss}
\end{figure}

The emergence of moir\'{e} pattern in twisted bilayers gives rise to a series of propagating diffracted beams with the in-plane wave vectors $|\bm g| < q_0$. To the lowest order in the permittivity modulation, there are four forward and four backward diffracted beams with the in-plane wave vectors $\pm\bm K$  and  $\pm\bm K'$, where $\bm K = \bm k_1 - \bm k_2$ and $\bm K' = \bm k_1' - \bm k_2'$. In the bilayers with small twist angles $\varphi \ll \omega/(ck)$, these beams slightly deviate from the bilayer normal. 

The diffracted beams are considerably polarized even for the unpolarized incident beam. Their polarization, in general, is elliptical, as sketched by green ellipses in Fig.~\ref{fig:Diff-Gauss}(a), which results from the bilayer twist. Typically, the beams have strong linear polarization degree along the corresponding diffraction wave vectors.

Figure~\ref{fig:Diff-Gauss}(b) shows the intensity of the forward diffracted beam as a function of the interlayer distance and light frequency.
The moir\'{e} diffraction stems from the near-field interaction and, therefore, is highly sensitive to the interlayer distance and the twist angle. The diffraction is most prominent at the waveguide resonances.   
The SAM and OAM of the diffracted beam are shown in Figs.~\ref{fig:Diff-Gauss}(c) and~\ref{fig:Diff-Gauss}(d). The diffracted beam acquires much larger spin and orbital angular momenta than the transmitted or reflected beams. This is especially noticeable for the SAM, Fig.~\ref{fig:Diff-Gauss}(c).

 In the case of small layer polarizability, $\alpha_0 k , \alpha_1 k \ll1$, the analytic expression for amplitudes of the moir\'e diffracted beams can be obtained, see Appendix~\ref{app:pert}. 
For small twist angles, $\varphi \ll 1$, the intensity of each forward diffracted beam is given by
\begin{equation}
    T^{(\bm K)}=\frac{q_0^2}{\varkappa^2}(2\pi \alpha_1)^4(\varkappa^4+q_0^4)[1+\cos(2 q_0 d)]\e^{-2\varkappa d}\;,
\end{equation}
while the SAM and OAM assume the form
\begin{align}
\label{eq:difr-S}
    &\langle s\rangle=\varphi\tan(q_0 d) \,\frac{ k^4}{\varkappa^4+q_0^4}\,, \\
\label{eq:difr-L}
&\langle l\rangle=-\varphi\tan(q_0 d)\, \frac{k^2\left(k^4-6k^2 q_0^2+4q_0^4\right)}{a^2 \varkappa^4 (\varkappa^4+q_0^4)}\;.
\end{align}
Both $\langle s \rangle$ and $\langle l \rangle$ in Eqs.~\eqref{eq:difr-S} and~\eqref{eq:difr-L}, respectively, are independent of the permittivity modulation $\alpha_1$. The reason is that the very occurrence of the moir\'e diffracted beams originates from the structure twist. The fluxes of the SAM and OAM as well as the photon flux are all proportional to $\alpha_1^4$, see Appendix~\ref{app:pert}. As a result,  the SAM and OAM per photon are independent of the permittivity modulation and can reach high values.

\subsection{Field distribution in interference zone}

\begin{figure}[t]
    \centering
    \includegraphics[width=0.99\columnwidth]{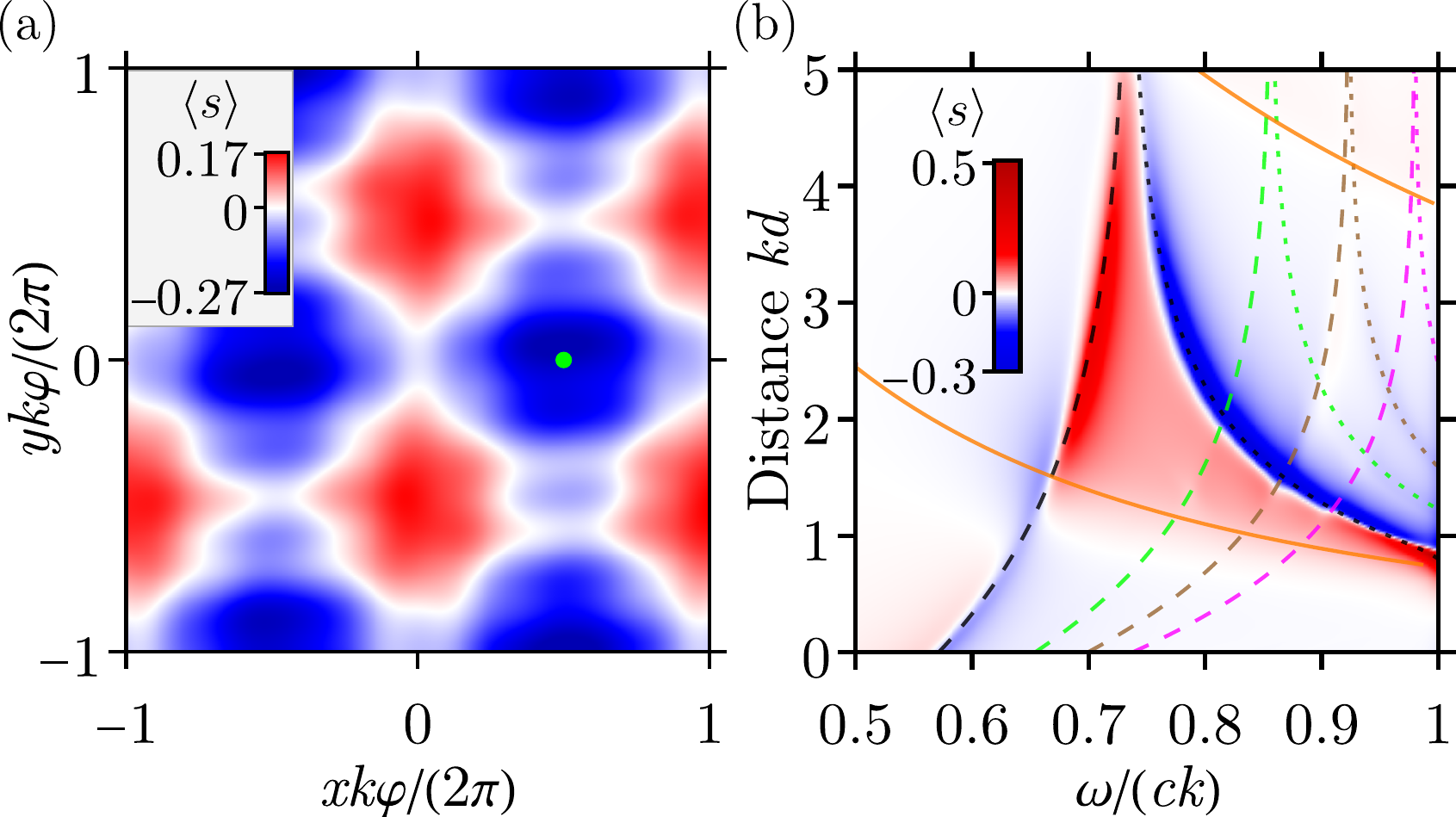}
    \caption{(a) The spatial distribution of the SAM behind the bilayer in the area of interference of transmitted and diffracted beams. The incident beam is linearly polarized, $\beta=\pi/4$, $\varphi=\pi/15$, $kd=2$, $\omega/(ck)=0.816$. Other parameters are listed in the caption to Fig.~\ref{fig:Trans-and-Refl}. (b) The SAM at the coordinate $\bm \rho=[\pi/(k\varphi),0]$, highlighted by the green point in Fig.~\ref{fig:Interfer}(a), as a function of the interlayer distance and the light frequency.} 
    \label{fig:Interfer}
\end{figure}

Finally, we consider the polarization of the field behind the bilayer in the far zone at $1/\varkappa \lesssim z-d \lesssim q_0/(\varphi^2 k^2)$. In this region,  the near field vanishes but the transmitted beam interferes with the forward diffracted beams. Thus, the total field reads
\begin{equation}
    \bm E(\bm \rho )=\bm{t}\bm E_{\rm in}+\sum_{\bm{g}=\pm \bm K, \pm \bm K'}\bm{t}^{(\bm g)}\bm E_{\rm in} \, \e^{\rmi  \bm g\cdot \bm \rho} \,.
\end{equation}

Now, we assume that the incident beam is linearly polarized at the angle $\beta$ to the $x \parallel \bm K'$ axis. Due to the interference, the total field acquires strong local circular polarization which oscillates in the plane, see Fig.~\ref{fig:Interfer}(a). The period of the oscillations $2\pi/(k\varphi)$ is determined by the moir\'{e} pattern. Figure~\ref{fig:Interfer}(b) shows the amplitude of the SAM spatial oscillations as a function of the interlayer distance and light frequency. The value of SAM at the waveguide resonances can be rather high.

 In the case of small layer polarizability, $\alpha_0 k , \alpha_1 k \ll1$, the spatial distribution of the SAM in the interference zone is given by
\begin{align}
    \langle s\rangle=&(4 \pi \alpha_1 k)^2\,\frac{q_0}{\varkappa}\,\cos(q_0 d)\, \e^{-\varkappa d}\\ \nonumber
    &\times\sin(2\beta)\left[\cos(k\varphi y)-\cos(k\varphi x)\right].
\end{align}
The spin polarization follows $\sin 2\beta$ dependence on the polarization direction of the incident light, as in a birefrigent medium. This local birefringence is induced by the moir\'{e} pattern and has maxima at the positions $[n\pi/(k\varphi), m\pi/(k\varphi)]$, where $n$ and $m$ are integers with different parity.

\section{Summary}\label{sec:sum}

To summarize, we have developed a theory of the spin and orbital chiral optical response of a twisted photonic bilayer consisting of dielectric slabs with in-plane modulation of permittivity.
Using the scattering formalism, we have calculated the Jones matrices relating the electric field of incident light with the electric fields of transmitted, reflected, and diffracted by the moir\'{e} pattern beams and studied the parameters of the beams as a function of the twist angle and the interlayer distance.

The interaction of light with twisted photonic bilayer includes the transfer of angular momentum. As a result, initially unpolarized beam
carrying no orbital angular momentum acquires spin angular momentum upon transmission and orbital angular momentum upon reflection.
The effects are particularly pronounced at the frequencies of the eigen optical modes of the twisted bilayer.
Both effects are sensitive to the interlayer distance but exhibit different behavior: the orbital angular momentum in reflection oscillates with a constant amplitude when the interlayer distance is increased while the spin angular momentum in transmission oscillates and decays exponentially  at large distances. 
The spin angular momentum emerges from the near-field interaction in the twisted bilayer and the related diffraction by the moir\'e pattern. The origin of the orbital angular momentum is the interference of the asymmetric distortions of the beam induced by the two layers.
The moir\'e diffracted beams are also found to carry significant spin and orbital angular momenta. The predicted effects can be observed in twisted photonic-crystal bilayers which are already available in the infrared~\cite{Lou2023} and microwave~\cite{Liu2022} spectral ranges.

\acknowledgments 

E.S.V. acknowledges the support by the Russian Science Foundation
(Project No. 23-12-00142). 

%%%%%%%%%%%%%%%%%%%%%%%%%%%%%%%%%%%%%%%
%%%%%%%%%%%%%%%%%%%%%%%%%%%%%%%%%%%%%%
%%%%%%%%%%%%%%%%%%%%%%%%%%%%%%%%%%%%%
%%%%%%%%%%%%%%%%%%%%%%%%%%%%%%%%%%%%
%%%%%%%%%%%%%%%%%%%%%%%%%%%%%%%%%%%
%%%%%%%%%%%%%%%%%%%%%%%%%%%%%%%%%%
%%%%%%%%%%%%%%%%%%%%%%%%%%%%%%%%%
%%%%%%%%%%%%%%%%%%%%%%%%%%%%%%%%

\appendix

\section{Solution of the Maxwell equations}\label{app:sol}

To calculate the electric field in the whole space 
we solve the wave equation 
\begin{equation}\label{eq:maxwell}
    \operatorname{rot}\operatorname{rot} \boldsymbol{E}-q_0^2\boldsymbol{E} = 4\pi q_0^2 [ \bm P^{(1)} \delta(z) + \bm P^{(2)} \delta(z-d) ]
\end{equation}
with the sheet polarizations $\bm P^{(j)}$ given by Eq.~\eqref{eq:Pj}.  
Substituting the solution in  the form of Eq.~\eqref{eq:Eall},
we obtain the amplitudes of the diffracted waves
\begin{align}\label{eq:Ez}
\bm E_{\bm g}'(\bm q_\parallel, z)  = 
\sum_j  \bm{\mathcal{D}}_{\boldsymbol q_\parallel + \bm g}  \boldsymbol{P}_{ \boldsymbol{g}}^{(j)}(\bm q_\parallel)\,e^{\rmi q_z(\bm g) |z-z_j|}  \,,
\end{align}
where
\begin{align}
\bm{\mathcal{D}}_{\boldsymbol q} = \frac{2\pi \rmi}{\sqrt{q_0^2 - q^2}} \left(q_0^2  - \bm q \otimes \bm q \right)
\end{align}
is the dyadic Green function and $q_z(\bm g) = \sqrt{q_0^2 - (\bm q_\parallel + \bm g)^2}$.
Note that the terms with $|\bm q_\parallel + \boldsymbol{g}| > q_0$ in the electric field distribution Eq.~\eqref{eq:Ez} correspond to evanescent waves. 

The Fourier harmonics of the polarization $\bm P^{(j)}_{\bm g}(\bm q_\parallel)$ in Eq.~\eqref{eq:Ez} are determined by the electric field in the layers,
 \begin{align}\label{eq:polarization}
  \boldsymbol{P}^{(j)}_{\bm g}(\bm q_\parallel) &=  \alpha_0 \boldsymbol{E}_{\boldsymbol{g}} (\bm q_\parallel,z_j) \\
 &+ \alpha_1 \sum_{n=\pm 1}
 [ \boldsymbol{E}_{\boldsymbol{g} + n \bm k_j} (\bm q_\parallel,z_j)  + \boldsymbol{E}_{\boldsymbol{g} + n \bm k'_j} (\bm q_\parallel,z_j)  ] \,, \nonumber \\
 \boldsymbol{E}_{\bm g} (\boldsymbol{q}_\parallel,z) &= \bm E_{\rm in}(\bm q_\parallel)  \e^{\rmi q_z z} \delta_{\bm g ,0} + \boldsymbol{E}'_{\bm g } (\boldsymbol{q}_\parallel,z) \,,
  \end{align}
which follows from Eq.~\eqref{eq:Pj}. We solve the set of Eqs.~\eqref{eq:Ez} and~\eqref{eq:polarization} numerically and, in some approximations, analytically. In numerical solution, to smoothen the sharp resonances we use the regularization
$\omega \to \omega + \rmi \gamma$ with small $\gamma$ which corresponds to the medium with weak losses.
 
The Jones matrices $\bm t^{(\bm g)}$ and $\bm r^{(\bm g)}$ for Fourier components of the diffracted beams propagating forward and backward are defined as
\begin{align}
\bm E_{ \bm g}(\bm q_\parallel ,d) &=  \bm t^{(\bm g)}(\bm q_\parallel) \bm E_{\rm in}({\bm q_\parallel })\,,\\
\bm E'_{ \bm g}(\bm q_\parallel , 0) &=   \bm r^{(\bm g)}(\bm q_\parallel) \bm E_{\rm in} ({\bm q_\parallel })\,.
\end{align}
In particular, $ \bm t(\bm q_\parallel) \equiv \bm t^{(0)}(\bm q_\parallel)$ and $ \bm r(\bm q_\parallel) \equiv  \bm r^{(0)}(\bm q_\parallel)$ determine the amplitudes of the transmitted and reflected beams.

%%%%%%%%%%%%%%%%%%%%%%%%%%%%%%%%%%%%%%%%%%

\section{Perturbative approach}\label{app:pert}

If the polarizabilities of the layers are small, $\alpha_0 k , \alpha_1 k \ll1$, they can be considered as a perturbation. Then, the electric field of the scattered waves can be obtained by an iterative solution of Eqs.~\eqref{eq:Ez}--\eqref{eq:polarization}.

To calculate the SAM of the transmitted beam, the transmission matrix $\bm t (0)$ up to fourth order in $\alpha_1$ shall be calculated. Indeed, for that the diffraction by both the first and the second layer should be taken into account, e.g., consequently by the reciprocal lattice vectors $\bm k_1$, $-\bm k_2$, $-\bm k_1$, $\bm k_2$. This yields the spin-dependent contribution to the transmission matrix
\begin{align}\label{eq:trans}
    &\delta \bm t (0)= \sum_{\substack{\bm g_1= \pm \bm k_1, \pm \bm k_1' \\ \bm g_2= \pm \bm k_2, \pm \bm k_2'}} (\bm M_{\bm g_2,\bm g_1} + \bm M_{\bm g_1,\bm g_2} \e^{2\rmi q_0 d}) ,
    \\
    & \bm M_{\bm g_2,\bm g_1} = \alpha_1^4 \e^{[\rmi q_z(\bm g_1- \bm g_2)-2\varkappa]d} \bm{\mathcal{D}}_0 \bm{\mathcal{D}}_{\bm g_2} \bm{\mathcal{D}}_{\bm g_1- \bm g_2} \bm{\mathcal{D}}_{\bm g_1}  ,
\end{align}
where $\varkappa=\sqrt{k^2-q_0^2}$. In the case of small twist angle $\varphi \ll 1$, the dominant contribution is given by  the terms with $\bm g_1 = \pm \bm k_1$, $\bm g_2 = \pm \bm k_2$ or $\bm g_1 = \pm \bm k_1'$, $\bm g_2 = \pm \bm k_2'$. Then, the result simplifies to
\begin{align}\label{eq:trans2}
    \delta \bm t (0)=2\alpha_1^4 \e^{\rmi q_0 d-2\varkappa d} &\bm{\mathcal{D}}^2_0
     \left\{\bm{\mathcal{D}}_{\boldsymbol{k}_2}  \bm{\mathcal{D}}_{\boldsymbol{k}_1}+\bm{\mathcal{D}}_{\boldsymbol{k}_2'}\bm{\mathcal{D}}_{\boldsymbol{k}_1'}\right.
     \\ \nonumber
     &\left. + \left[ \bm{\mathcal{D}}_{\boldsymbol{k}_1}  \bm{\mathcal{D}}_{\boldsymbol{k}_2}+\bm{\mathcal{D}}_{\boldsymbol{k}_1'}\bm{\mathcal{D}}_{\boldsymbol{k}_2'} \right]\e^{2\rmi q_0 d}\right\} \,.
\end{align}
For the case of non-polarized excitation, the SAM of the transmitted wave is calculated as
\begin{align}
\langle s \rangle = \frac12 {\rm Tr\,}[ \e^{-\rmi q_0 d}\bm \sigma_2 \,\delta \bm t (0) + \delta \bm t^\dag (0)\, \bm \sigma_2\, \e^{\rmi q_0 d}] ,
\end{align} 
which yields Eq.~\eqref{eq:st}.

%%%%%%%%%%%%%%%%%%%

To calculate the OAM for the reflected beam, it is sufficient to take into account the second-order diffraction processes. However, the finite width of the beam $a$ and the dependence  of the reflection amplitude on the in-plane wave vector $\bm q_\parallel$ should be considered. 
The second-order contribution to the reflection matrix reads
\begin{align}\label{eq:dr}
    \delta \bm{r}&(\bm q_\parallel) =\alpha_1^2 \bm{\mathcal{D}}_{\bm q_\parallel} \\ \nonumber
    &\times
     \Big( \sum_{\bm g_1 = \pm \bm k_1, \pm \bm k_1'} \bm{\mathcal{D}}_{\bm q_\parallel + \bm g_1} 
     +\e^{2\rmi q_0 d} \sum_{\bm g_2 = \pm \bm k_2, \pm \bm k_2'} \bm{\mathcal{D}}_{\bm q_\parallel + \bm g_2} 
    \Big) \,.
\end{align}
Now we expand $\bm{r}(\bm q_\parallel)$ to the second order in $\bm q_{\parallel}$ as follows
\begin{align}
\bm{r}(\bm q_\parallel) = r +\frac12 \delta\bm{r}_{xx}'' q_x^2 + \delta\bm{r}_{xy}'' q_x q_y+ \frac12 \delta\bm{r}_{yy}'' q_y^2\;,
\end{align}
where we use the fact that the linear-in-$\bm q_{\parallel}$ terms vanish. 
Then, the OAM is readily expressed via the expansion coefficients
\begin{align}
    \langle l\rangle=&-\frac{1}{4a^4|r|^2} \Im {\rm Tr}\left[\delta\bm{r}_{xy}''^{\dagger} (\delta\bm{r}_{yy}''- \delta\bm{r}_{xx}'') \right].
\end{align}
Substituting the explicit expression from Eq.~\eqref{eq:dr}, we arrive to Eq.~\eqref{eq:lt}

%%%%%%%%%%%%%%%%%%%%%%%%%%%%%%%%%%%%%%%%%%%%%%%

The amplitude of the beams diffracted by the moir\'{e} wave vector $\bm K = \bm k_1 - \bm k_2$ is also obtained in the second order and reads
\begin{equation} \label{eq:difr-E}
    \bm{t}^{(\bm K)}(\bm q_\parallel)=\alpha_1^2\e^{-\varkappa d}\bm{\mathcal{D}}_{\bm K +\bm q_\parallel}\left[ \bm{\mathcal{D}}_{\bm q_\parallel+\boldsymbol{k}_1}+ \bm{\mathcal{D}}_{\bm q_\parallel-\boldsymbol{k}_2} \e^{2 \rmi q_0 d}\right]\,.
\end{equation}
We expand $\bm{t}^{(\bm K)}(\bm q_\parallel)$ to the first order in $\bm q_\parallel$ as follows
\begin{align}
\bm{t}^{(\bm K)}(\bm q_\parallel) = \bm{t}^{(\bm K)} + \bm{t}^{(\bm K)\prime}_x q_{x} + \bm{t}^{(\bm K)\prime}_y q_{y} .
\end{align}
Then, the SAM and OAM are given by
\begin{align}
&\langle s \rangle  = \frac{ {\rm Tr}[\bm{t}^{(\bm K)\dag} \bm  \sigma_2\,  \bm{t}^{(\bm K)}]}{ {\rm Tr}[\bm{t}^{(\bm K)\dag}   \bm{t}^{(\bm K)}]}, 
& \langle l\rangle=\frac{\Im{\rm Tr}[\bm{t}^{(\bm K)\prime\dag}_x \bm{t}^{(\bm K)\prime}_y  ]}{ a^2 {\rm Tr}[\bm{t}^{(\bm K)\dag}   \bm{t}^{(\bm K)}]}\;,
\end{align}
 respectively.
 \bibliography{AM}

\end{document}